\documentclass[12pt,twoside]{article}
\usepackage{fleqn,espcrc1}
\usepackage{latexsym} 
\usepackage{amssymb}  
\usepackage{amsfonts} 
\usepackage{multirow}
\usepackage{epsf}
\usepackage{graphicx}
\def\y0{y^{(0)}}

\newcommand \beq{\begin{eqnarray}}
\newcommand \eeq{\end{eqnarray}}

\newcommand{\mnote}[1]{\marginpar{\tiny {}}}   

\begin{document}

\title{\bf Towards the Quark-Gluon Plasma
}
\author{
   Peter Braun-Munzinger\\
   Gesellschaft f{\"u}r Schwerionenforschung\\
   64220 Darmstadt, Germany\\
   }
\maketitle

\begin{abstract}

    \noindent We discuss recent experimental results in the field of
    ultra-relativistic nuclear collisions. The emerging ``picture'' is
    a collectively expanding, initially  hot and dense fireball in which
    strangeness- and low-mass di-lepton pair production are enhanced
    and J/$\Psi$ production is suppressed compared to expectations
    from nucleon-nucleon collisions. It is argued that, taken
    together, these data provide circumstantial evidence that a (at
    least partly) partonic phase was produced in such collisions.

\end{abstract}


\section{Introduction} 
\noindent Research with ultra-relativistic nuclear collisions aims at
producing, in the laboratory, quark-gluon plasma. This new state of
matter is predicted to exist at high temperatures and/or high baryon
densities. Specifically, numerical solution of QCD using lattice
techniques imply that the critical temperature (at zero baryon
density) is about 170 MeV \cite{lattice}. Comprehensive surveys of the
various experimental approaches of how to produce such matter in
nucleus-nucleus collisions have been given recently
\cite{jsinpc,pbmqm97,bass}.  Here we focus on a few very selected
aspects, namely radial flow, strangeness production, dilepton
production, and J$\Psi$ measurements and explore possible correlations
among these observations. The aim is to elucidate possible connections
of these observations to the QCD phase transition.

\section{Radial Flow} 
\begin{figure}[thb]

\vspace{-1cm}

\epsfxsize=11cm
\begin{center}
\hspace*{0in}
\epsffile{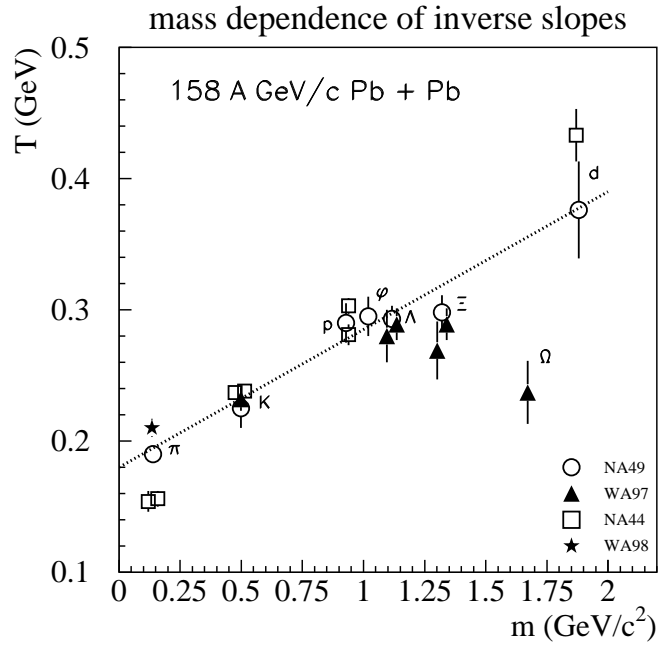}
\end{center}

\vspace{-2.5cm}

\caption{
Inverse slope parameters vs particle mass.
} 
\label{fig:t_vs_m}
\end{figure}

\noindent A large body of data on transverse momentum (or transverse
mass) distributions of hadrons demonstrates that the inverse slope
constant of these (in general exponential) spectra scales linearly
with particle mass m. The experimental facts for Pb+Pb collisions at SPS
energy have been compiled recently \cite{jsinpc} and are shown in
Fig. ~\ref{fig:t_vs_m}. The observed linear relationship is naturally
interpreted in terms of a collectively expanding fireball, where $p_t
=p_t^{thermal}+ m\gamma_t\beta_t$ \cite{jsinpc,pbmqm97} with $\beta_t$
the transverse flow velocity and $\gamma_t
=\sqrt{1/(1-\beta_t^2)}$. Using the known relationship between $<p_t>$
and T,m one may deduce a mean value of the transverse velocity of
$<\beta_t> \approx 0.4$, in good agreement with results from
hydrodynamic descriptions \cite{pbmqm97}. However, the T-m relation
shown in Fig. ~\ref{fig:t_vs_m} also implies T=180 MeV for m=0, not
consistent with the current interpretation that thermal freeze-out
takes place at around T=120 MeV \cite{pbmqm97}.  

The now detailed
measurements of transverse momentum distributions also imply little
centrality dependence of the observed slope constants, in conflict
with an interpretation of these observations in terms of initial state
scattering \cite{pbmqm97}.

An interesting and somewhat puzzling deviation from the general trend
is observed for  multi-strange baryons: the corresponding slope
parameters are significantly smaller than expected. This has been
interpreted as evidence for early freeze-out \cite{vanhecke}. It could
be due to very small cross sections for elastic pion-strange baryon
scattering. 

The overall conclusion, however, is that the picture of a collectively
expanding fireball has survived all tests of the past years.

\section{Strangeness Enhancement and Equilibration}
\begin{figure}[thb]

\vspace{-1cm}

\epsfxsize=11cm
\begin{center}
\hspace*{0in}
\epsffile{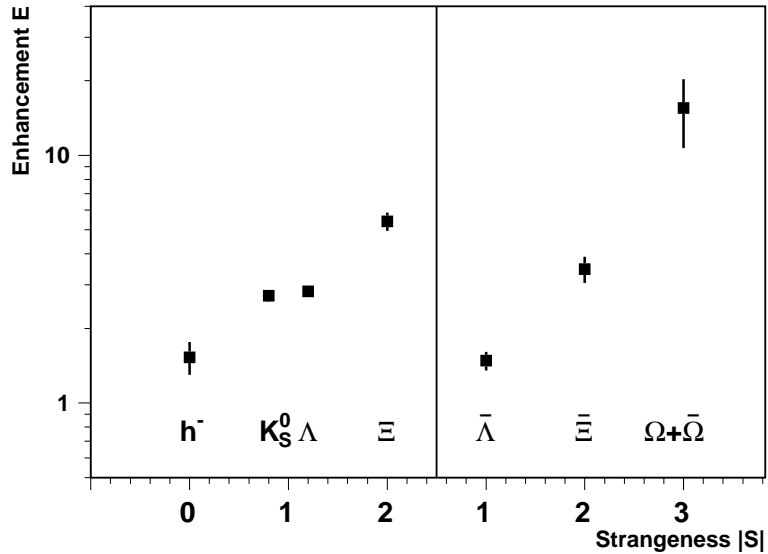}
\end{center}

\vspace{-1.5cm}

\caption{
Strangeness enhancement in Pb+Pb collisions for multi-strange
baryons. The data are from the WA97 collaboration \cite{wa97}.
} 
\label{fig:strange}
\end{figure}

\noindent The production yields of strange hadrons are significantly
increased in ultra-relativistic nuclear collisions compared to what is
expected from a superposition of nucleon-nucleon collisions. This has
been observed by several experiments both at the AGS and at the
SPS. To demonstrate the degree of enhancement observed we show, in
Fig. ~\ref{fig:strange}, the results of the WA97 collaboration for
multi-strange baryons \cite{wa97}.

The observed enhancement of more than one order of magnitude can
currently not be understood within any of the hadronic event
generators\footnote{For a discussion see the proceedings of the Quark
Matter 99 conference, Nucl. Phys. A, to be published.}. Surprizingly,
it is quantitatively explained if one assumes complete chemical
equilibrium in the hadronic phase of the collision
\cite{therm3}. Similar observations have been made for analyses of
S-induced SPS data \cite{therm2} and for AGS data \cite{therm1}.

\begin{figure}[thb]

\vspace{-1.7cm}

\epsfxsize=11cm
\begin{center}
\hspace*{0in}
\epsffile{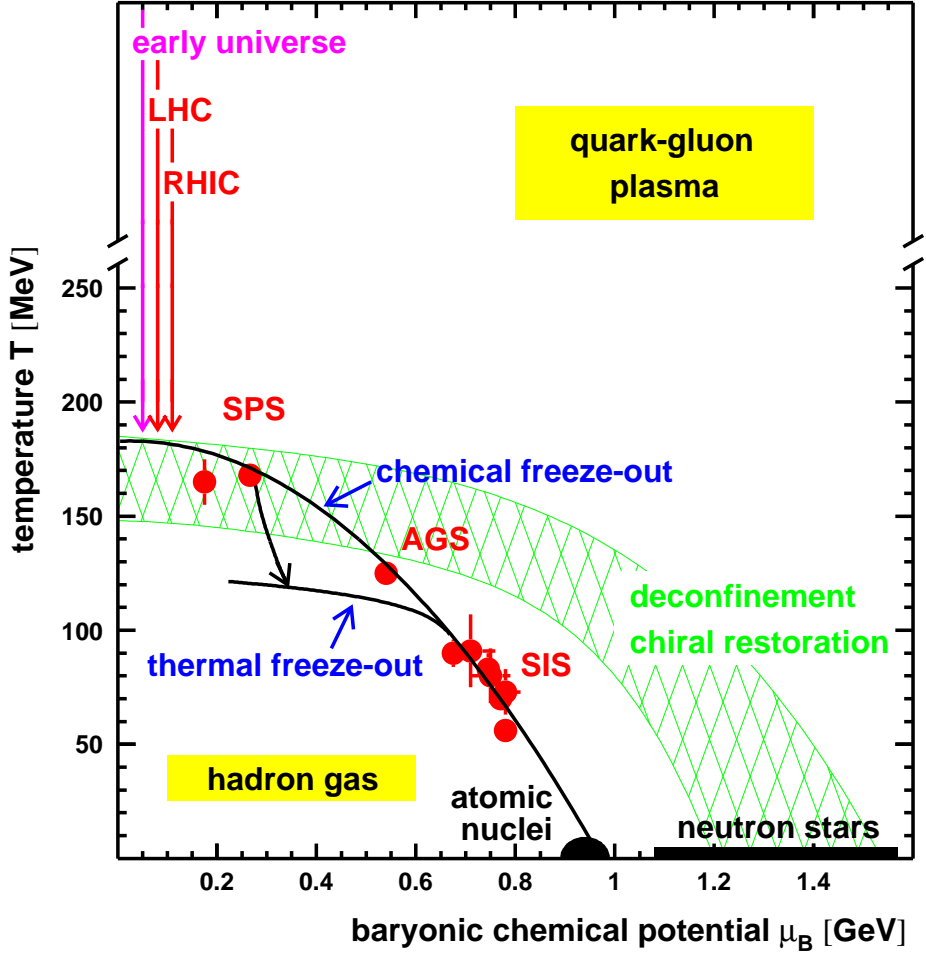}
\end{center}

\vspace{-2.0cm}

\caption{
Phase diagram of hadronic and partonic matter. The hadrochemical
freeze-out points are determined from thermal model analyses of heavy
ion collision data at SIS, AGS and SPS energy. The hatched region
indicates the current expectation for the phase boundary based on lattice QCD
calculations at $\mu$=0. The arrow from chemical to thermal
freeze-out for the SPS corresponds to isentropic expansion.
} 
\label{fig:phase-d}
\end{figure}

How chemical equilibration can be reached in a purely hadronic
collision is not clear in view of the small production cross section
for strange and especially multi-strange hadrons. In fact, system
lifetimes of the order of 50 fm/c or more are needed for a hot
hadronic system to reach full chemical equilibration
\cite{life}. Such lifetimes are at variance with lifetime values
established from interferometry analyses, where upper limits of about
10 fm/c are deduced \cite{appels}. 

Another very interesting observation is that the chemical
potentials $\mu$ and temperatures T resulting from the thermal analyses of
\cite{therm3,therm2,therm1} place the systems at chemical freeze-out
very close to where we currently believe is the phase boundary between
plasma and hadrons. This is demonstrated in Fig. ~\ref{fig:phase-d}
\footnote{This is an updated version of the figure shown in
\cite{jsinpc,pbmqm97}.}  where also results from lower energy analyses
are plotted. The freeze-out trajectory (solid curve through the data
points) is just to guide the eye but follows closely the empirical
curve of \cite{freeze}.

The closeness of the freeze-out parameters (T,$\mu$) to the phase
boundary might be the clue to the apparent chemical equilibration in
the hadronic phase: if the system prior to reaching freeze-out was in
the partonic (plasma) phase, then strangeness production is determined
by larger partonic cross sections as well as by hadronization. Slow
cooking in the hadronic phase is then not needed to produce the
observed large abundances of strange hadrons. Early simulations of
strangeness production in the plasma and during hadronization support
this interpretation at least qualitatively \cite{knoll}. Faced with
the present results a new theoretical look seems mandatory.

\section{Enhancement of Low Mass Dilepton Pairs}
\begin{figure}[thb]

\vspace{-3.0cm}

\epsfxsize=11cm
\begin{center}
\hspace*{0in}
\epsffile{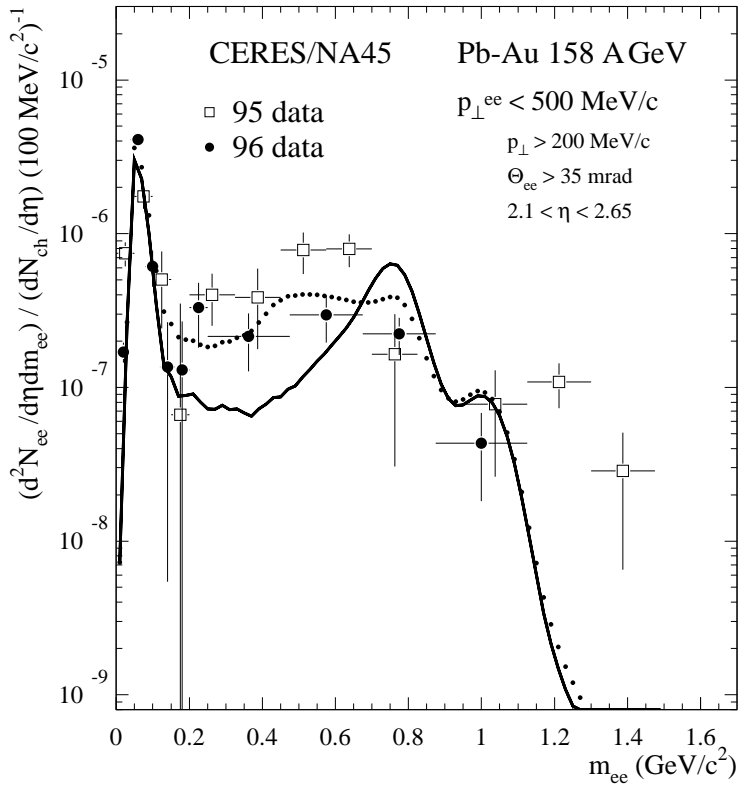}
\end{center}

\vspace{-0.80cm}

\caption{ Invariant mass distribution of electron pairs for central
Pb-Au collisions and low pair transverse momentum. The solid line
represents the expectation for the yield based on extrapolating
nucleon-nucleon collision results. The dashed line is obtained by
taking medium modifications of the $\rho$ meson into account \cite{rapp}  }
\label{fig:cer}
\end{figure}

\noindent The CERES and HELIOS/3 collaborations found that, in central
nucleus-nucleus collisions at SPS energy, low mass (m $<$ 800 MeV)
dilepton pairs are produced at yields which are significantly larger
than expected from nucleon-nucleon collisions
\cite{ceres1,ceres2,helios}.  The enhancement is concentrated at low
pair transverse momentum \cite{lenkeit} as can be seen from the recent
CERES data which are presented in Fig. ~\ref{fig:cer}.

The observed enhancement has been attributed to changes of the mass
and/or width of the $\rho$ meson in the hot and dense fireball. The
still somewhat controversial situation has been reviewed recently
\cite{rw}.  Here we want to add two points. First, the enhancement
sets in at centralities corresponding to less than 35 \% of the total
inelastic cross section \cite{lenkeit}. This implies impact parameters
b $<$ 8 fm (see Fig. 2 in \cite{pbmqm97}. From thereon it scales
quadratically with particle multiplicity. Secondly, the $\rho$ mesons
are formed according to Fig. ~\ref{fig:phase-d} at around T=165
MeV. These two facts suggest that pion number (i.e. temperature) and
$\pi-\pi$ collisions in the medium are at the origin of the observed
enhancement.

\section{J/$\Psi$ Suppression}
\noindent 
The suppression of J/$\Psi$ mesons (compared to what is expected from
hard scattering models) was early on predicted \cite{matsui} to be a
signature for color deconfinement. Data for S-induced collisions
exhibited a significant suppression but systematic studies soon
revealed that such suppression exists already in p-nucleus collisions
and is due to the absorption in (normal) nuclear matter of a color
singlet $c \bar c g$ state that is formed on the way towards J/$\Psi$
production. The situation has been summarized in
\cite{lourenco,kharzeev,jsinpc}.

\begin{figure}[thb]

\vspace{-3.5cm}

\epsfxsize=18cm
\begin{center}
\hspace*{0in}
\epsffile{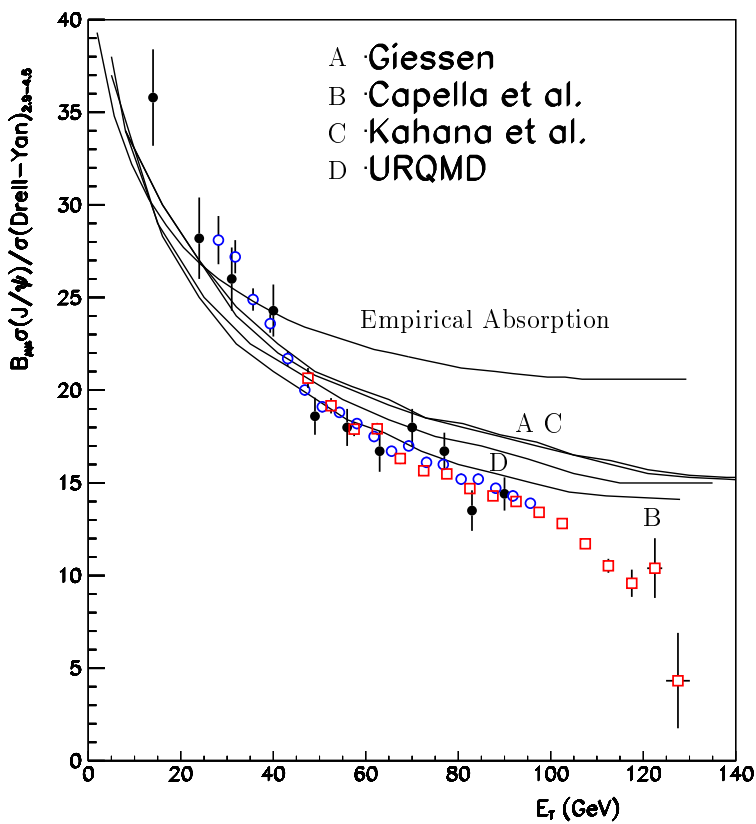}
\end{center}

\vspace{-11.0cm}

\caption{Transverse energy dependence of J/$\Psi$ production in Pb+Pb
collisions normalized to the Drell-Yan cross section. Data and the
empirical absorption curve are from the Na50 collaboration
\cite{na50,cicalo}. The various theoretical distributions summarize
summarize attempts to describe the observations in ``conventional''
scenarios. For details see text. } 
\label{fig:jpsi}
\end{figure}

The data for Pb+Pb collisions now exhibit clear evidence for anomalous
absorption beyond the standard absorption expected for such
systems. The most recent results are summarized in
Fig. ~\ref{fig:jpsi}, taken from \cite{na50,cicalo}. There seems to be
a break away from the standard absorption curve at around a transverse
energy value of 40 GeV corresponding to an impact parameter of about 8
fm \cite{na50}. We note that this impact parameter value agrees with
the value from where on anomalous dilepton enhancement is observed by
the CERES collaboration (see above)\footnote{The data from CERES do
not reach lower centralities corresponding to larger impact
parameters. Hence this point is not yet ``water-tight''.}. The new
minimum bias data from NA50 (open points in Fig. ~\ref{fig:jpsi})
with their much smaller error bars very much accentuate the difference
to the standard absorption curve.

Whether these data provide unambiguous evidence  for the existence of
a deconfined phase in central Pb+Pb collisions is hotly
debated. However, there is at present no convincing explanation of the
observed data in standard scenarios without plasma. The theoretical
curves in Fig. ~\ref{fig:jpsi} show this. All calculations from the
Giessen group \cite{giessen}, from the Kahana team \cite{kahana}, from
Capella's group \cite{capella}, and from the Frankfurt group using
UrQMD \cite{urqmd} are based on models for the destruction of
charmonium by comoving pions, strings, etc. The corresponding
dissociation cross sections are poorly known \cite{pbm_d}. Despite
very different assumptions about these cross sections and despite a
number of other nontrivial assumptions (see, e.g. the discussion in
\cite{bmuller}) none of the calculations reproduce the data. For
example, comparing the Giessen calculation with the 1996 data (solid
points in Fig. ~\ref{fig:jpsi}) yields a reduced $\chi^2$ of larger
than 4, while comparison of any of the calculations with the high
statistics minimum bias data (open points in  Fig. ~\ref{fig:jpsi})
yields reduced $\chi^2$ values of larger (sometimes much larger) than 10.
We conclude that the charmonium suppression observed by the NA50
collaboration in central Pb+Pb collisions is highly non-trivial.

\section{Summary and Outlook}
\noindent Taken together, the observations of flow, strangeness
enhancement, enhancement of low-mass dilepton pairs, and charmonium
suppression lend strong support to the interpretation that, during the
course of a central Pb+Pb collision at SPS energy, an at least partly
deconfined state of matter, i.e. of quark-gluon plasma, has been
created. Two further rounds of experiments at the SPS will provide the
possibility to consolidate this picture. 

Meanwhile, experiments at the RHIC collider are about to commence and
the planning for the ALICE experiment at the LHC is well
underway. Physics prospects for experiments at these accelerators are
bright \cite{pbmqm99,gyulassy99}. Extrapolating from the results of  the AGS
and SPS program we expect much higher energy densities and
temperatures at collider energy. Production and study of a deconfined
phase over a large space-time volume should be possible.


\begin{thebibliography}{9}
\vspace{3mm}

\bibitem{lattice} F. Karsch, Proc. Lattice99, Nucl. Phys. B (in
print), hep-lat/9909006.

\bibitem{jsinpc} J. Stachel, Proc. INPC, Paris, August 1998,
Nucl. Phys. {\bf A654} (1999) 119c, nucl-ex/9903007.

\bibitem{pbmqm97} P. Braun-Munzinger and J. Stachel, Nucl. Phys. {\bf A638}
(1998) 3c.

\bibitem{bass} S. Bass, M. Gyulassy, H. St\"ocker, W. Greiner,
J.Phys. {\bf G25} (1999) R1-R57.

\bibitem{vanhecke} H. van Hecke, H. Sorge, N. Xu,
Phys. Rev. Lett. {\bf 81} (1998) 5764.

\bibitem{wa97} E. Andersen et al., Wa97 Coll., J. Phys. G:
    Nucl. Part. Phys. {\bf 25} (1999) 171, E. Andersen, et al., WA97
    Coll., Phys.  Lett. {\bf B449} (1999) 401.


\bibitem{therm3} P. Braun-Munzinger, I. Heppe, J. Stachel, 
nucl-th/9903010,  Phys. Lett B(in print).


\bibitem{therm2} P. Braun-Munzinger, J. Stachel, J. P. Wessels,
N. Xu, Phys. Lett. B {\bf 365} (1996) 1.  


\bibitem{therm1}P. Braun-Munzinger, J. Stachel, J. P. Wessels, N. Xu,
Phys. Lett. B {\bf 344} (1995) 43.  

\bibitem{life} J. Sollfrank and U. Heinz in: Quark Gluon Plasma 2,
R.C. Hwa, editor, World Scientific 1996, p. 555.

\bibitem{appels} H. Appelsh\"auser et al., NA49
collaboration, Eur. Phys. J. {\bf C2} (1998) 661.

\bibitem{freeze} J. Cleymans and K. Redlich, nucl-th/9903063. 

\bibitem{knoll} H. W. Barz, B. L. Friman, J. Knoll, H. Schulz,
Nucl. Phys. {\bf A519} (1990) 831.


\bibitem{ceres1} G. Agakichiev et al., CERES collaboration,
Phys. Rev. Lett. {\bf 75} (1995) 1272.

\bibitem{ceres2} G. Agakichiev et al., CERES collaboration, Phys. Lett. {\bf
B422} (1998) 405.

\bibitem{helios} M. Masera for the HELIOS collaboration,
Nucl. Phys. {\bf A590} (1995) 93c.
 
\bibitem{lenkeit} B. Lenkeit for the CERES collaboration, Proc. Quark
Matter 99 conference, Torino, June 1999, Nucl. Phys. A(in print).

\bibitem{rapp} R. Rapp, proceedings Rencontres de Moriond '98,
nucl-th/9804065; R. Rapp and C. Gale, hep-ph/9902268.

\bibitem{rw} R. Rapp and J. Wambach, Adv. Nucl. Phys. (in print),
nucl-th/9909229.  


\bibitem{matsui} T. Matsui and H. Satz, Phys. Lett. {\bf B178} (1986) 416.


\bibitem{lourenco} C. Lourenco, Nucl. Phys. {\bf A610} (1996) 552c.

\bibitem{kharzeev} D. Kharzeev, Nucl. Phys. {\bf A638} (1998) 279c.


\bibitem{na50} M.C. Abreu et al., NA50 collaboration, Phys. Lett. {\bf
B450} (1999) 456.

\bibitem{cicalo} C. Cicalo for the NA50 collaboration, Proc. Quark
Matter 99 conference, Torino, June 1999, Nucl. Phys. A(in print).

\bibitem{giessen} J. Geiss, C. Greiner, E.L. Bratkovskaya, W. Cassing,
U. Mosel, Phys. Lett. {\bf B447} (1999) 31; J. Geiss, C. Greiner,
E.L. Bratkovskaya, W. Cassing, nucl-th/9810059.
\bibitem{kahana} D.E. Kahana, S.H. Kahana,
Prog. Part. Nucl. Phys. {\bf 42} (1999) 395;
D.E. Kahana, S.H. Kahana, nucl-th/9908063.

\bibitem{capella} N. Armesto, A. Capella, E.G. Ferrero,
Phys. Rev. {\bf C59} (1999) 395.

\bibitem{urqmd} C. Spieles, R. Vogt, L. Gerland, S.A. Bass,
M. Bleicher, H. St\"ocker, W. Greiner, hep-ph/9902337.

\bibitem{pbm_d} P. Braun-Munzinger and K. Redlich, nucl-th/9908026.

\bibitem{bmuller} B. M\"uller,  Proc. Quark
Matter 99 conference, Torino, June 1999, Nucl. Phys. A(in print),
nucl-th/9906029. 

\bibitem{pbmqm99} P. Braun-Munzinger, Proc. Quark
Matter 99 conference, Torino, June 1999, Nucl. Phys. A(in print),
nucl-exp/9908007.

\bibitem{gyulassy99} S. Bass et al., Proc. Quark
Matter 99 conference, Torino, June 1999, Nucl. Phys. A(in print),
nucl-th/9907090.


\end{thebibliography}
\end{document}